\documentclass[11pt]{article}
\usepackage{amsmath,amssymb, graphicx,euscript,url,bm}
\usepackage{physics}
\usepackage{bm}
\usepackage{mathrsfs}
\usepackage{mathtools}
\usepackage{cite}
\usepackage{siunitx}
\usepackage{float}
\usepackage[font=small,skip=0pt]{caption}
\usepackage[caption = false]{subfig}
\usepackage{slashed}
\allowdisplaybreaks

\begin{document}

\vspace{0.2cm}
\begin{center}
\Large\bf\boldmath
Quasielastic Lepton-Nucleus Scattering and the Correlated Fermi Gas Model
\end{center}

\begin{center}
{\sc Sam Carey}\\

{\it 
 Department of Physics and Astronomy, \\
Wayne State University, Detroit, Michigan 48201, USA}

\end{center}
\begin{center}
\large\boldmath
Contribution to the 25th International Workshop on Neutrinos from Accelerators
\end{center}

\vspace{0.2cm}
\begin{abstract}
  \vspace{0.2cm}
  \noindent
The study of neutrino-nucleus scattering processes is important for the new generation neutrino experiments for better understanding of the neutrino oscillation phenomenon. A significant source of uncertainty in the cross-section comes from limitations in our knowledge of nucleon and nuclear effects in the scattering process. Here we present a fully analytic implementation of the Correlated Fermi Gas nuclear model for the quasi-elastic lepton nucleus scattering. The implementation is then used to separately compare form factor effects and nuclear model effects, specifically the Relativistic Fermi Gas nuclear model, for both electron-carbon and neutrino-carbon scattering data.
\end{abstract}
\vspace{0.2cm}

\section{Introduction}

The charge-current quasi-elastic (CCQE) scattering cross-section of charged leptons off a nucleus involves folding the lepton-quark interaction twice, at the nucleon and nuclear levels. These are described in terms of nucleon form factors and the nuclear model, which contribute as two major sources of systematic uncertainty. This work, based on our recent study in \cite{Bhattacharya:2024}, discusses the importance of controlling the uncertainties arising from the nucleon and nuclear levels. This is essential for precision measurements of Standard Model (SM) parameters and for studying non-standard neutrino interactions (NSI).

We present an analytical implementation of the Correlated Fermi Gas (CFG) nuclear model \cite{Hen:2014yfa}, characterized by a depleted Fermi gas region and a high-momentum tail region incorporating short-range correlated nucleon pairs. This model is then compared to the widely known Relativistic Fermi Gas (RFG) model using electron and neutrino scattering data on carbon, with a fixed nucleon form factor. Comparisons are also made across various form factors keeping the nuclear model fixed.

\section{Lepton Nucleus Scattering: CFG Implementation}
\begin{figure}[H]
\begin{center}
\includegraphics[scale=0.37]{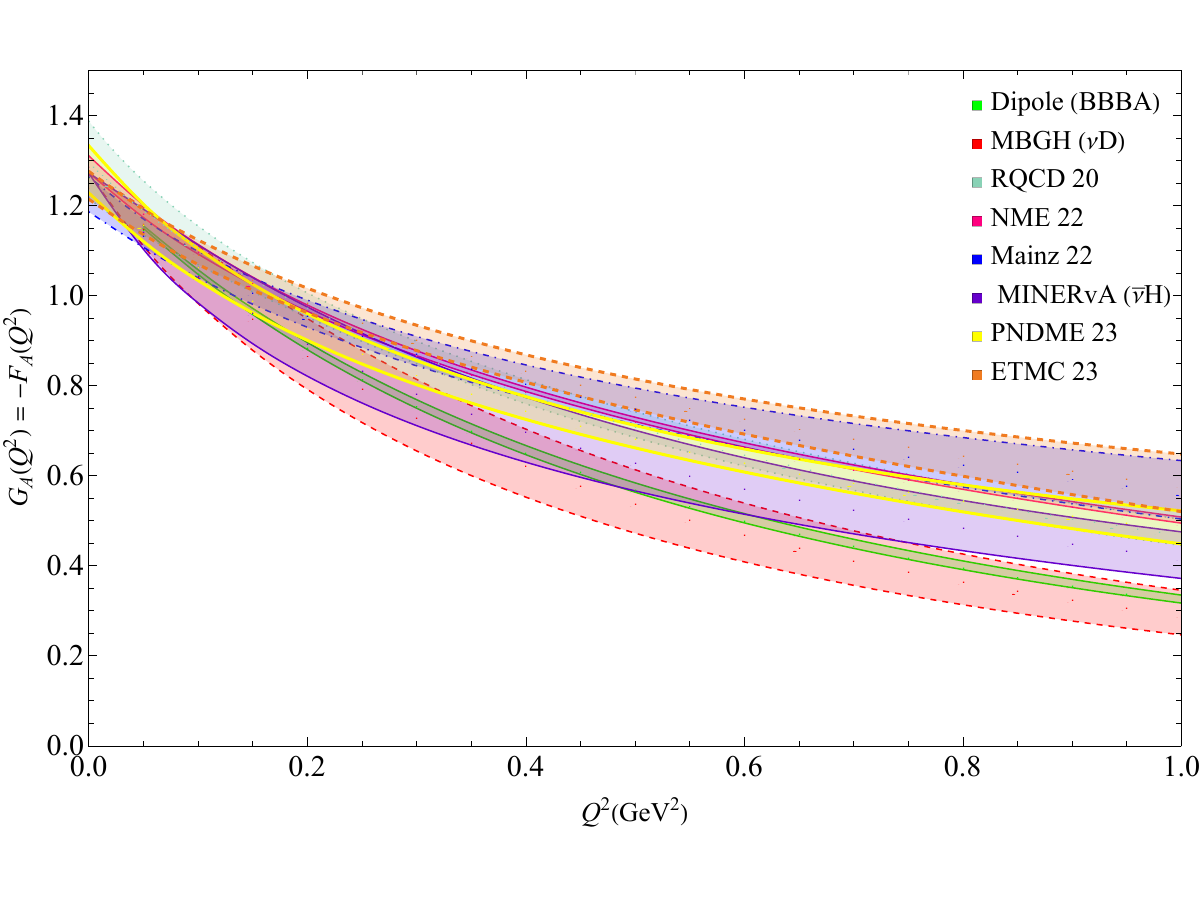}
\hspace{1cm}
\includegraphics[scale=0.268]{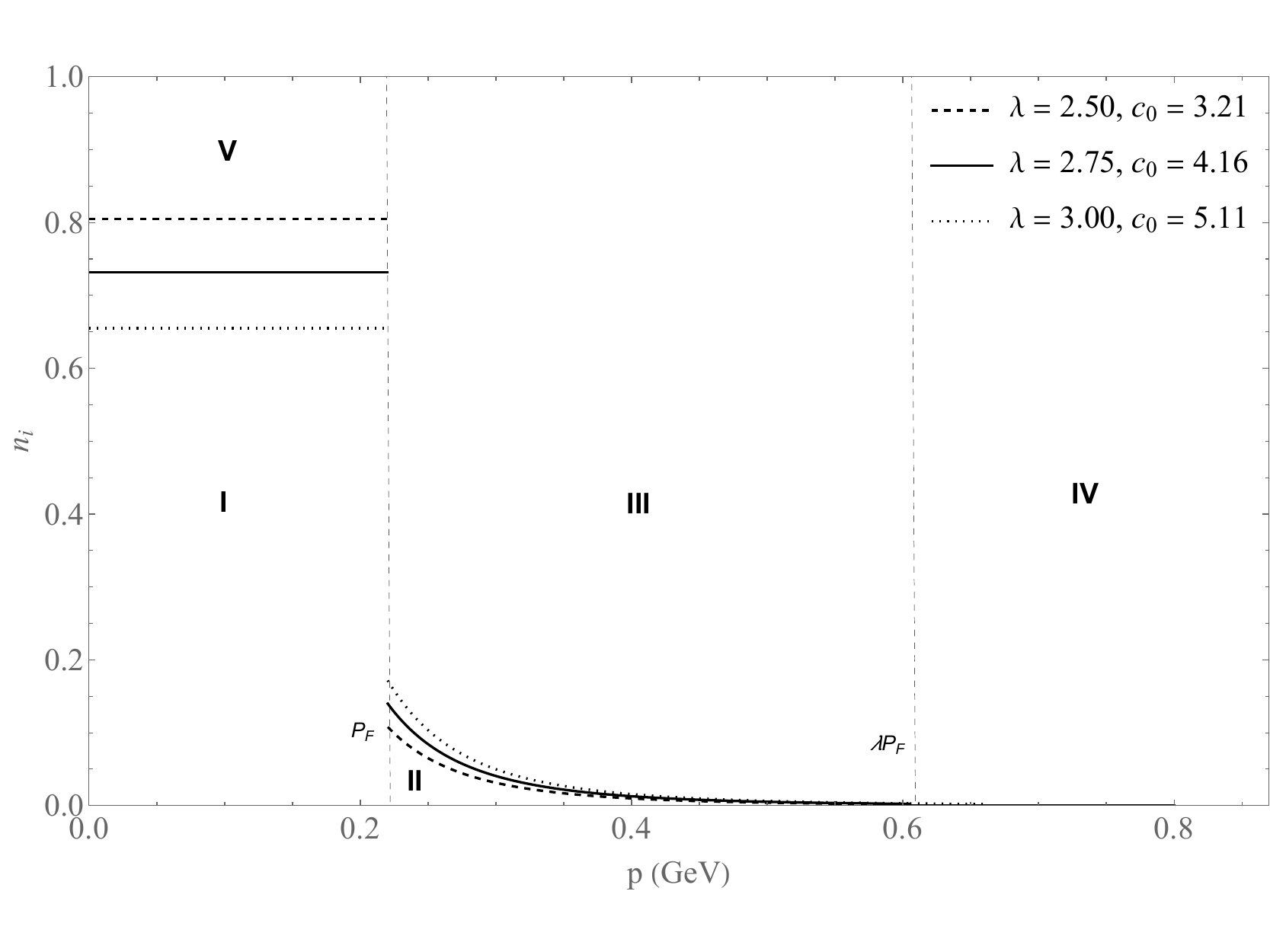}
\caption{\label{Fig:1} \textbf{Left}: the comparison between the different axial form factor parameterizations: Dipole (BBBA) using $M_A = 1.014\pm0.014$ GeV \cite{Bradford:2006yz}, MBGH \cite{Meyer:2016oeg}, RQCD 20 \cite{RQCD:2019jai}, NME 22 \cite{Park:2021ypf}, Mainz 22 \cite{Djukanovic:2022wru}, MINERvA \cite{MINERvA:2023avz}, PNDME 23 \cite{Jang:2023zts} and ETMC 23 \cite{Alexandrou:2023qbg} \textbf{Right}: The distribution of nucleons in the CFG model with the different lines
corresponding to the extremal variation of $\lambda$ and $c_0$.}
\end{center}
\end{figure}

The differential cross-section for an incoming lepton with energy $E_\ell$ scattering off a nucleus at an angle $\theta_\ell$ with respect to the outgoing lepton is given by, 
\begin{eqnarray}
\frac{d\sigma^{\,\ell}_{\rm nuclear}}{dE_\ell\, d\cos\theta_\ell} &\propto& \mathcal{C}^2_{EM,Weak} L^{\mu\nu}W_{\mu\nu}  \\
\mathcal{C}_{EM, Weak} &=& \frac{4\pi\alpha}{q^2}, \,\frac{G_F}{\sqrt{2}}\,|V_{ij}| \nonumber
\end{eqnarray}
where $\alpha$ is the fine structure constant, $q$ the 4-momentum transfer, $G_F$ the Fermi coupling constant, $|V_{ij}|$ the CKM matrix element for flavor $ij$, $L^{\mu\nu}$ the Lepton tensor and $W_{\mu\nu}$ the nuclear tensor described in terms of the nuclear distribution and hadron tensor as,
\begin{eqnarray} 
W_{\mu\nu} =\mathcal{N}\int d^3{p}\, n_i(\bm{p})  [1-n_f(\bm{p}+\bm{q})]  H_{\mu\nu}(\epsilon_{\bm{p}}, \bm{p}; q^0, \bm{q} ) \frac{\delta( \epsilon_{\bm{p}} - \epsilon^\prime_{\bm{p}+\bm{q}} + q^0 )}{\epsilon_{\bm{p}} \epsilon_{\bm{p}+\bm{q}}},
\end{eqnarray}
where $ p = (\epsilon_{\bm{p}}, \bm{p})$,  $ p^\prime = (\epsilon_{\bm{p+q}}, \bm{p+q})$ are the momenta of the initial and final state nucleon respectively and $\mathcal{N}$ is a normalization factor.
The hadron tensor is decomposed into scalar terms with coefficients expressed in terms of the nucleon form factor. For the electron scattering involving vector form factors we implement the dipole based BBBA \cite{Bradford:2006yz} and the $z$-expansion based BHLT form factors \cite{Borah:2020gte}. In the case of neutrino scattering an additional axial form factor based on $z$-expansion for the nucleon is included. The different parametrizations for the axial form factor are shown in the left side of Figure \ref{Fig:1}. In the RFG model the initial nucleons are distributed in the region with $p\leq p_F$ (denoted I in the right plot of Figure \ref{Fig:1}) and the final nucleons in the region with $p > p_F$ (denoted III and IV in the right plot of Figure \ref{Fig:1}). In the CFG model, the nucleons are distributed in regions I, II initially and post-interaction can avail the phase space in regions III, IV and V as depicted in the right side of Figure \ref{Fig:1}. The total cross section is then calculated by summing the contributions from the six possible transitions. More detailed description of the implementation is discussed in \cite{Bhattacharya:2024}.
\section{Results}
\begin{figure}[H]
\begin{center}
\includegraphics[scale=0.38]{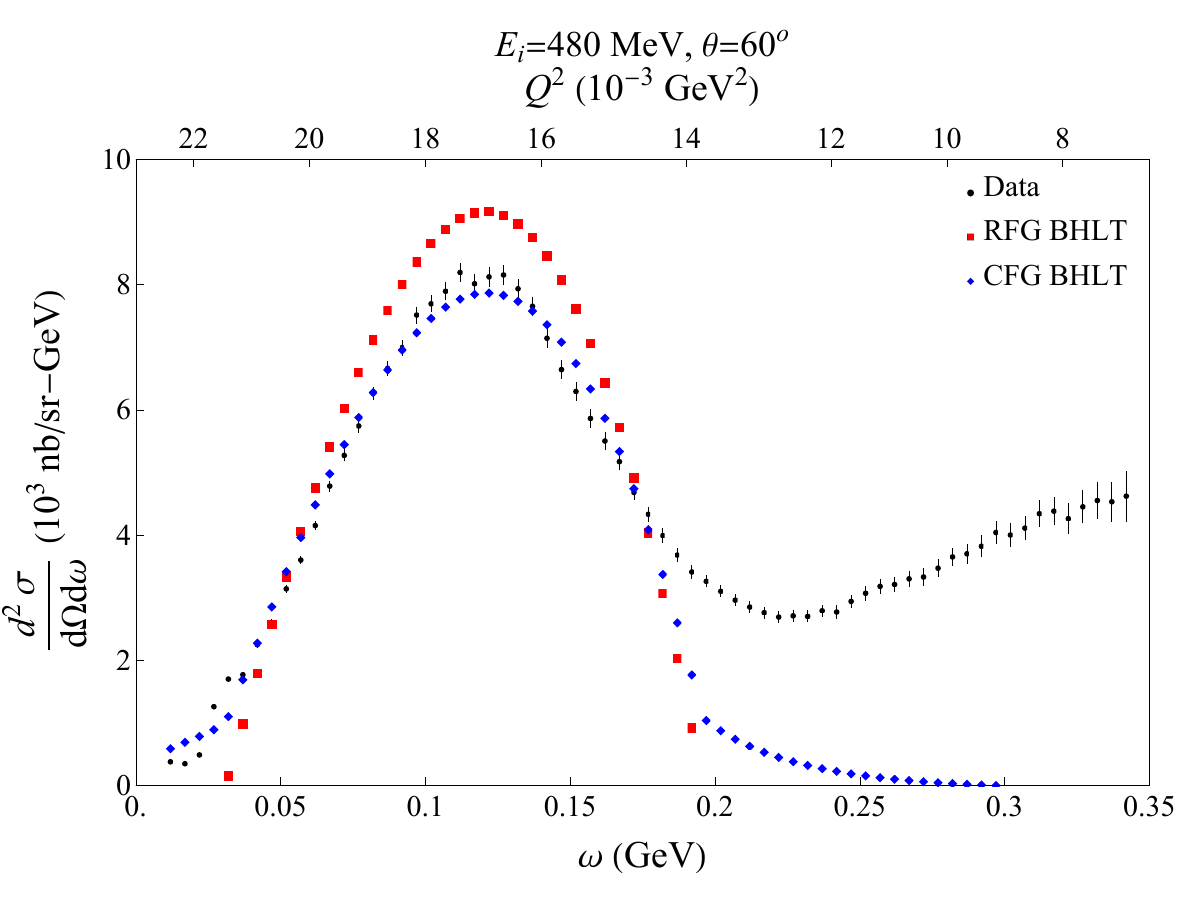}
\hspace{1cm}
\includegraphics[scale=0.38]{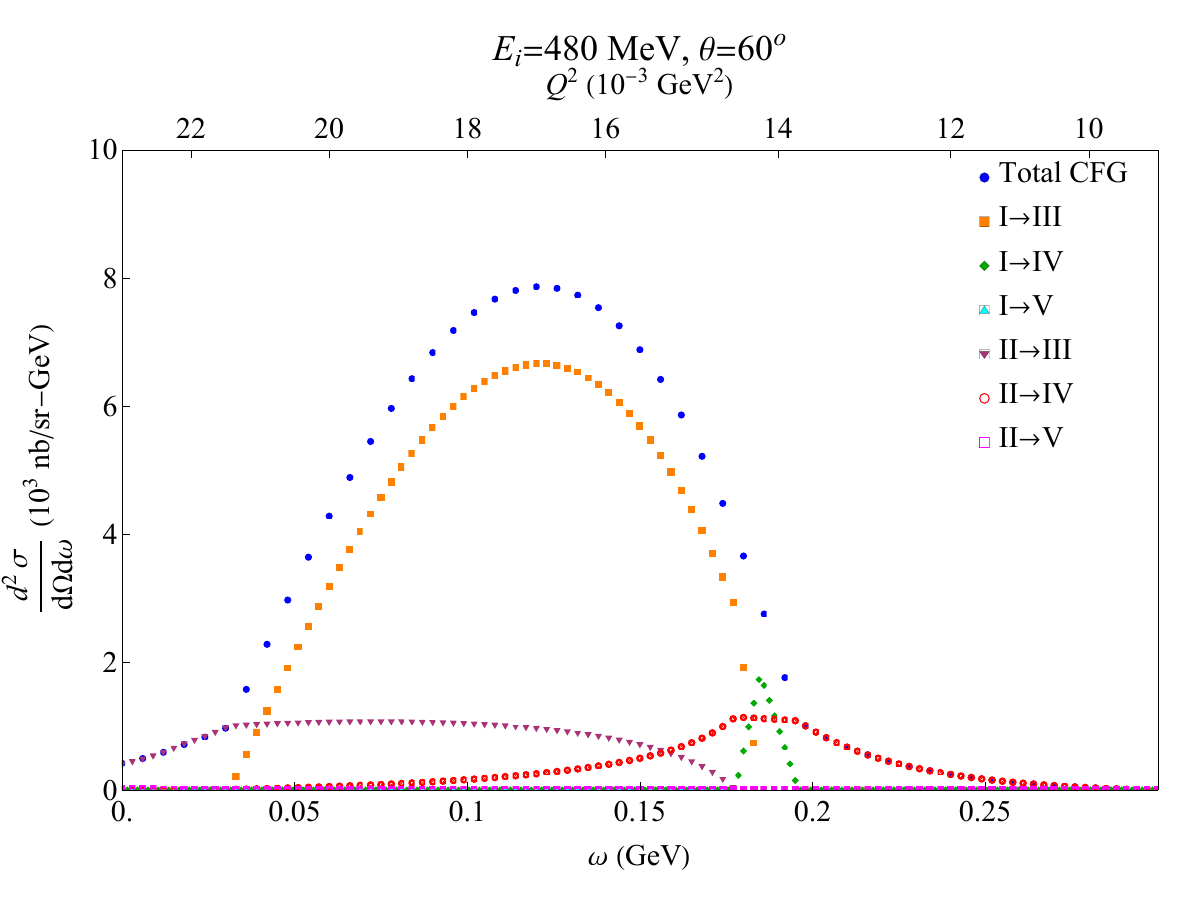}
\caption{\label{Fig:2} \textbf{Left}: A comparison of the RFG (red) and CFG (blue) nuclear model with the Carbon scattering data. \textbf{Right}: The contribution to the CFG implementation from various transitions. The incoming electron energy is $480$ MeV and scattering off at $60^\circ$.} 
\end{center}
\end{figure}

\begin{figure}[H]
\begin{center}
\includegraphics[scale=0.38]{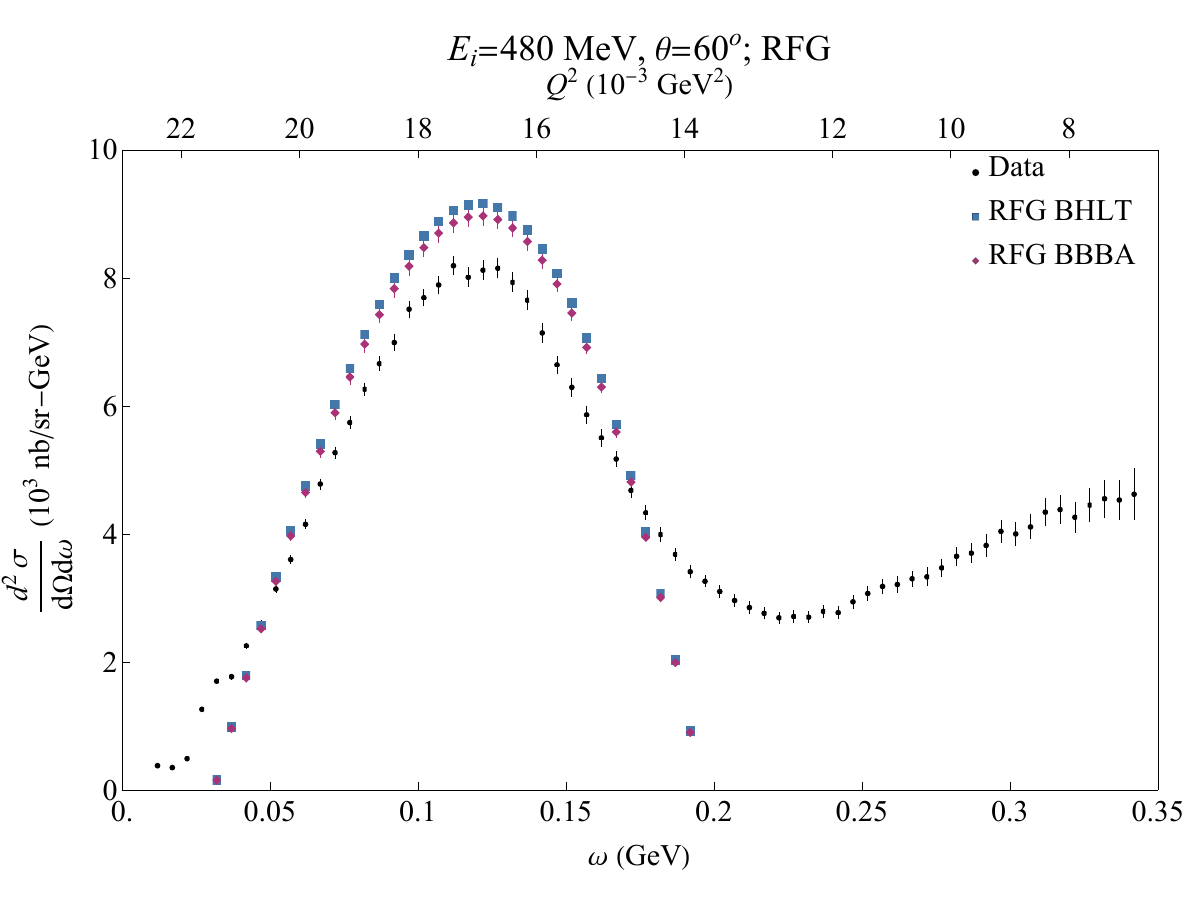}
\hspace{1cm}
\includegraphics[scale=0.38]{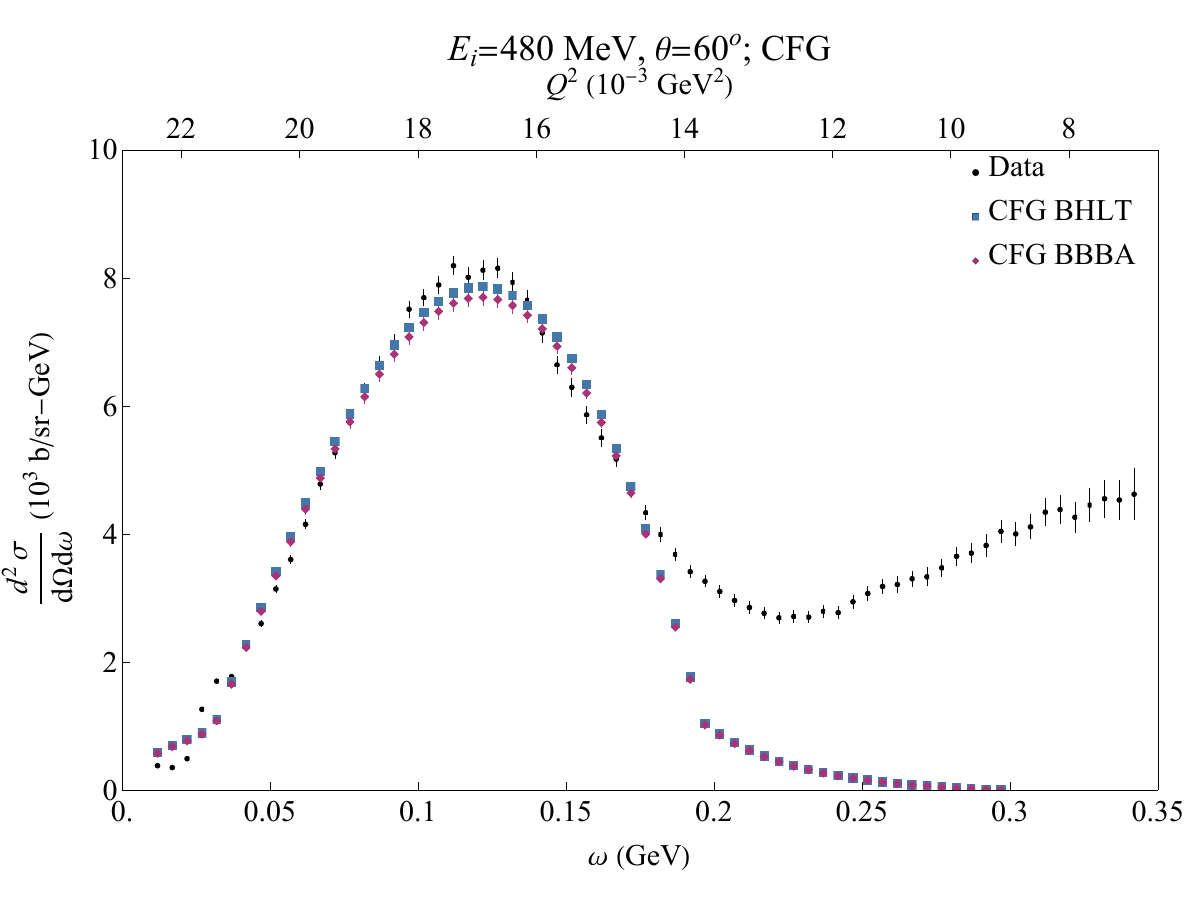}
\caption{\label{Fig:3} A comparison of the BHLT (teal) and BBBA (purple) vector form factor with the Carbon scattering data while keeping the \textbf{Left}: RFG nuclear model fixed, \textbf{Right}: CFG nuclear model fixed. The incoming electron energy is $480$ MeV and scattering off at $60^\circ$.}
\end{center}
\end{figure}

\begin{figure}[H]
\begin{center}
\includegraphics[scale=0.38]{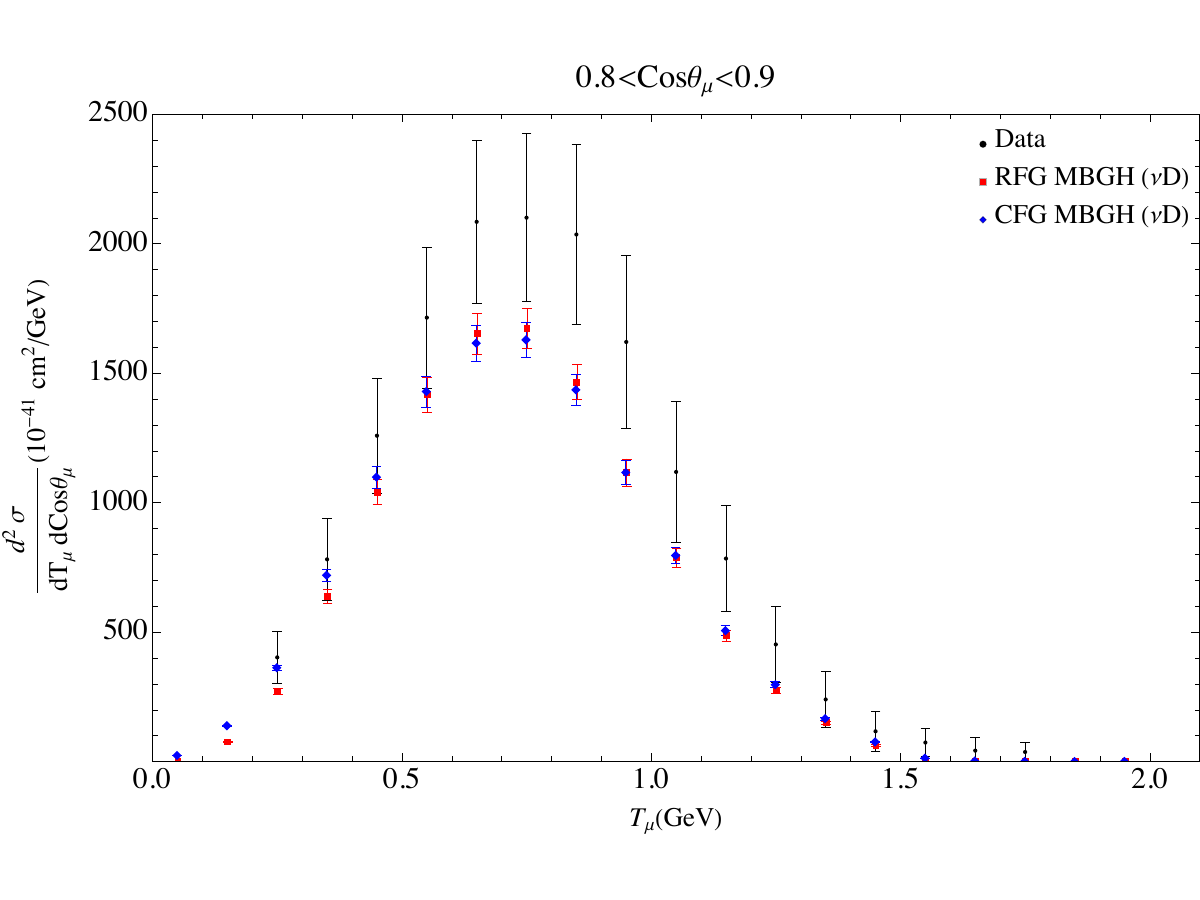}
\hspace{1cm}
\includegraphics[scale=0.38]{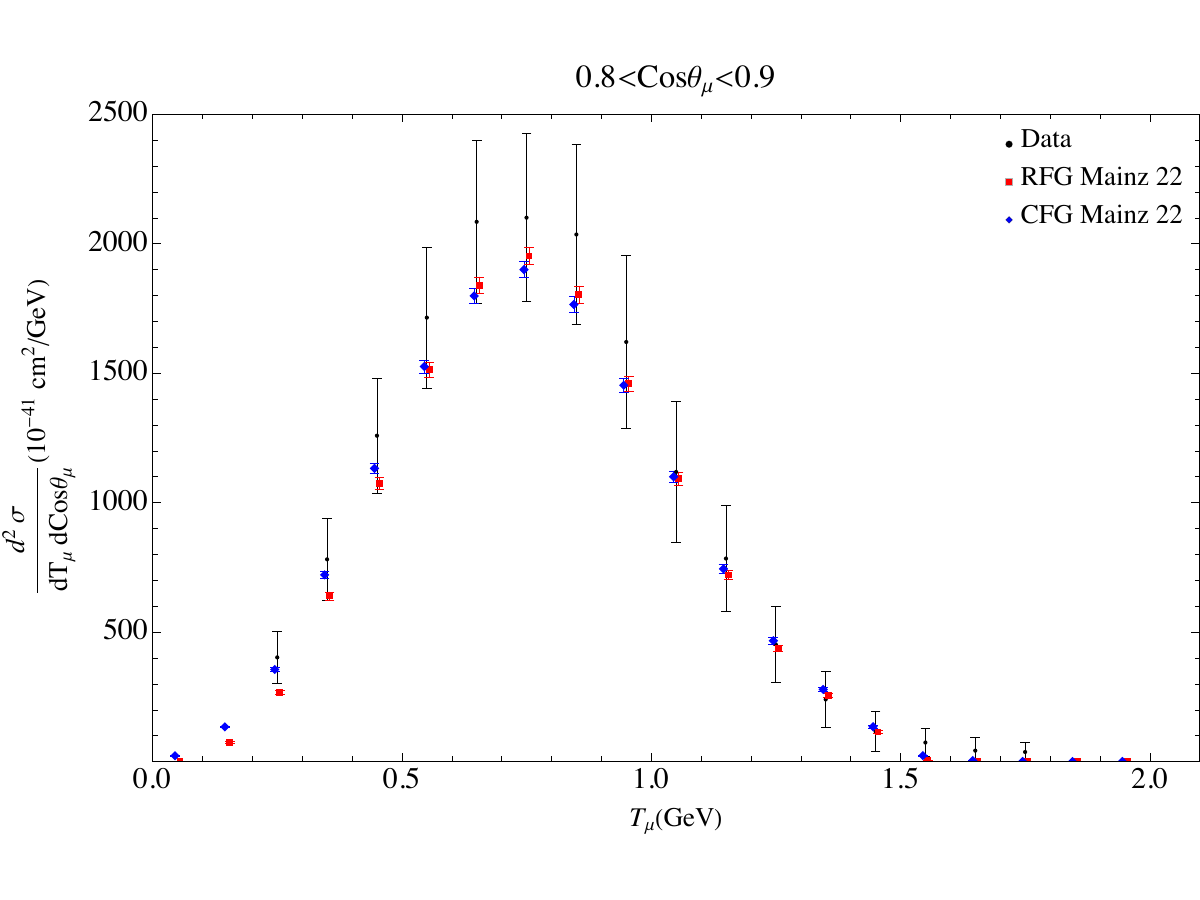}
\caption{\label{Fig:4} A comparison of RFG (red) and CFG (blue) models of flux averaged neutrino - carbon scattering data as a function of the outgoing muon kinetic energy scattering at an angle of $0.8 < \cos\theta_\mu<0.9$ using \textbf{Left}: MBGH axial form factor \textbf{Right}: Mainz 22 for the axial form factor.}
\end{center}
\end{figure}

\begin{figure}[H]
\begin{center}
\includegraphics[scale=0.38]{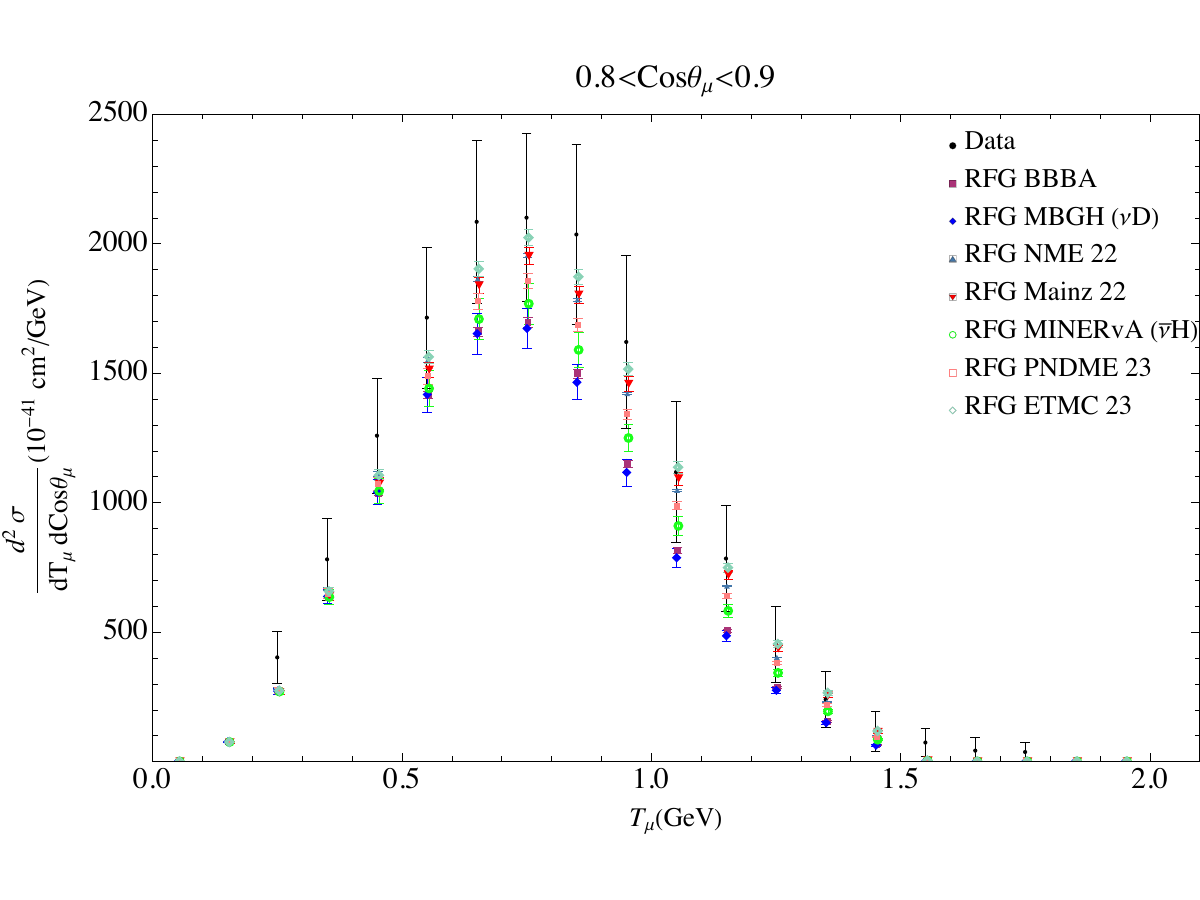}
\hspace{1cm}
\includegraphics[scale=0.38]{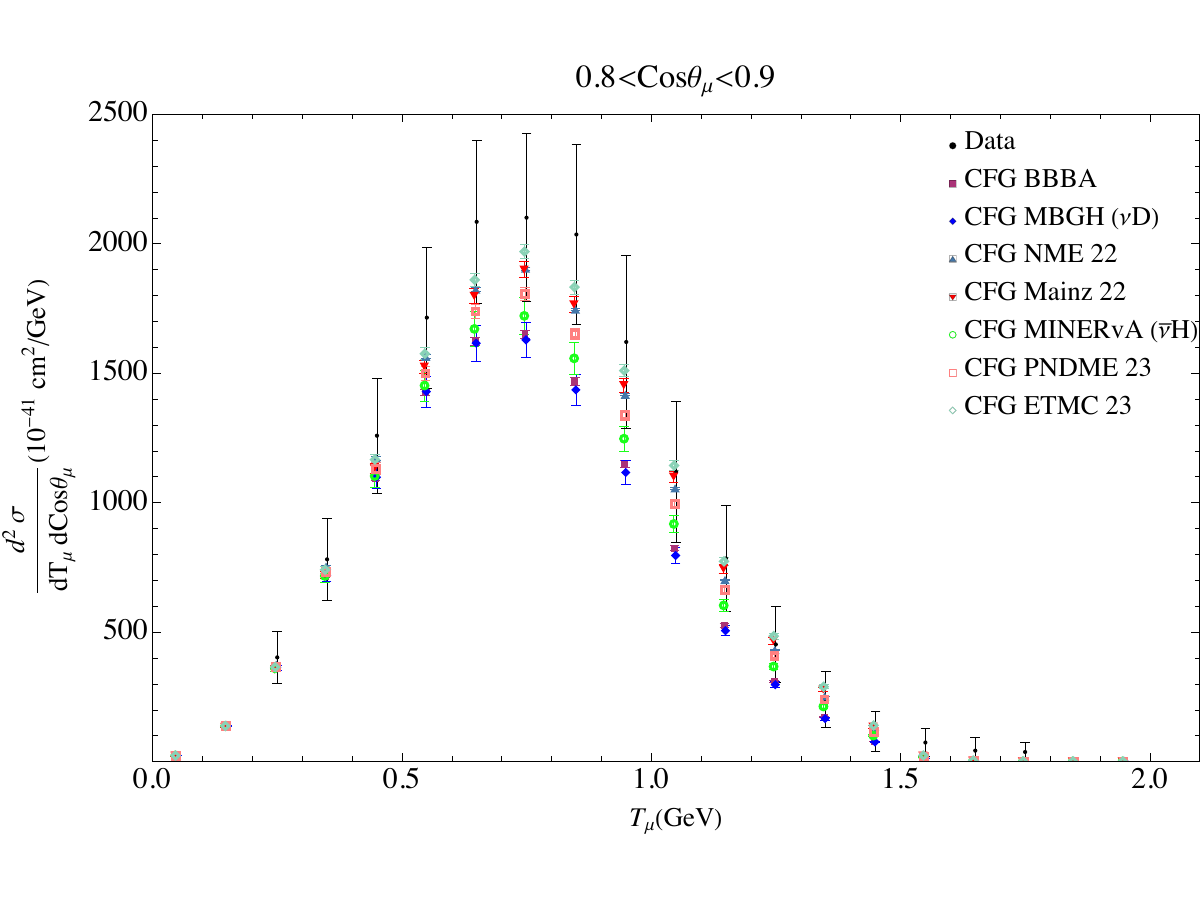}
\caption{\label{Fig:5} A comparison of BBBA, MBGH, NME22, Mainz22, MINER$\nu$A, PNDME 23 and ETMC 23 form factor parametrization to flux averaged neutrino - carbon scattering data as a function of the outgoing muon kinetic energy scattering at an angle of $0.8 < \cos\theta_\mu<0.9$ using \textbf{Left}: RFG model \textbf{Right}: CFG model.}
\end{center}
\end{figure}

First, we compare cross-section between the two nuclear models for \textbf{electron scattering} off carbon, while keeping the form factor fixed. This is shown for an incoming electron with energy $480$ MeV and scattered off at $60^\circ$ in comparison with the Carbon data \cite{Barreau:1983} on the left side of Figure \ref{Fig:2}. We observe a similar trend of the difference between the nuclear models for other kinematics as well. The right side of Figure \ref{Fig:2} depicts the underlying contributions to various transitions for this particular kinematic. 
In Figure \ref{Fig:3} we now compare the cross-sections between the two vector form factors while keeping the nuclear model fixed. It can be seen that difference from the form factors is small compared to difference due to the nuclear models.
\indent For \textbf{neutrino scattering} we compute the flux averaged cross-section and compare it to the MiniBooNE data \cite{MiniBooNE:2010bsu}. First, we fix the axial form factor and vary the nuclear model. Unlike in electron scattering, due to the flux averaging the differences here are very small as seen in Figure \ref{Fig:4}. But when the nuclear model is fixed and the axial form factor is varied, a continuous spread in the cross-section can be seen as in Figure \ref{Fig:5}.
\section{Conclusion and Future Outlook}
In this talk we presented an analytic implementation of the Correlated Fermi Gas (CFG) model for quasi-elastic electron-nucleus and neutrino-nucleus scattering. This implementation was then used to separate form factors and nuclear model effects. The CFG model compared to the RFG model, differed by the redistribution of strength from peak to the tail region. The CFG model shows better agreement with data for specific kinematics in electron scattering. However, distinguishing between nuclear models for flux-averaged neutrino cross sections remains a challenge. In the future it would be interesting to study the implementation of CFG model for semi-inclusive observables along with the incorporation of final-state interactions.
\section*{Acknowledgments} I would like to thank G. Paz and B. Bhattacharya for careful reading of the manuscript. This work was supported by the U.S. Department of Energy grant DE - SC0007983.
  \bibliography{CCQE}
\bibliographystyle{hunsrt}

\end{document}